\documentclass[aps,pra,preprint,showpacs,showkeys,superscriptaddress]{revtex4-1}
\usepackage{pstricks}
\usepackage{epsfig}
\usepackage{graphicx}
\usepackage{amsmath}
\usepackage{amsfonts}
\usepackage{amssymb}
\usepackage{wasysym}
\usepackage{bbold}
\usepackage{rotating}
\newcommand{\bra}[1]{\langle #1|}
\newcommand{\ket}[1]{|#1\rangle}

\begin{document}
\title{A primal-dual semidefinite programming algorithm tailored to the variational determination of the two-body density matrix}
\author{Brecht Verstichel}
\email{brecht.verstichel@ugent.be}
\affiliation{Ghent University, Center for Molecular Modeling, Technologiepark 903, 9052 Zwijnaarde, Belgium}
\author{Helen van Aggelen}
\affiliation{Ghent University, Department of Inorganic and Physical Chemistry, Krijgslaan 281 (S3), B-9000 Gent, Belgium}
\author{Dimitri Van Neck}
\affiliation{Ghent University, Center for Molecular Modeling, Technologiepark 903, 9052 Zwijnaarde, Belgium}
\author{Patrick Bultinck}
\affiliation{Ghent University, Department of Inorganic and Physical Chemistry, Krijgslaan 281 (S3), B-9000 Gent, Belgium}
\author{Stijn De Baerdemacker}
\affiliation{Department of Physics and Astronomy, Proeftuinstraat 86, 9000 Gent, Belgium}
\begin{abstract}
The quantum many-body problem can be rephrased as a variational determination of the two-body reduced density matrix, subject to a set of $N$-representability constraints. The mathematical problem has the form of a semidefinite program. We adapt a standard primal-dual interior point algorithm in order to exploit the specific structure of the physical problem. In particular the matrix-vector product can be calculated very efficiently. We have applied the proposed algorithm to a pairing-type Hamiltonian and studied the computational aspects of the method. The standard $N$-representability conditions perform very well for this problem.
\end{abstract}
\maketitle
\section{Introduction}
It was realized in the 1950's \cite{husimi,lowdin} that the energy of a quantum many-body system can be expressed in terms of the two-body reduced density matrix (2DM), when only one- and two-body interactions are present. This insight led to the idea of variationally determining the 2DM by minimizing the energy, henceforth referred to as the v2DM method. Once the 2DM is known, all other physical properties that can be expressed as one- or two-body operators can be extracted. In this way the 2DM effectively replaces the wave function and we have "quantum mechanics without wave functions" \cite{coleman_book}. Early attempts, however, produced unrealistic results \cite{mayer} and it was soon realized \cite{tredgold} that non-trivial constraints are needed to ensure that the 2DM is derivable from a physical wave function. These constraints were called $N$-representability conditions by Coleman \cite{coleman}, and Garrod and Percus \cite{garrod} derived two such conditions, the so-called $Q$ and $G$ conditions, which can be expressed as matrix-positivity constraints. With these constraints there were some attempts, some of which quite successful, to solve this problem numerically in the 1970s \cite{fusco,garrod_comp,rosina,mihailovic}. However the method was soon abandoned because of the computational cost.
Interest in the subject was renewed at the beginning of this century, when first Nakata \cite{nakata_first} and then Mazziotti \cite{mazziotti} realized that the v2DM problem can be formulated as a semidefinite program (SDP) for which general-purpose primal-dual SDP solvers can be used \cite{vandenberghe}, and they calculated the ground-state properties of small atoms and molecules. Primal-dual interior point methods are the "Rolls Royce" of SDP algorithms, having several appealing features, but they require a lot of storage and are computationally expensive. These early calculations were therefore limited to small systems (minimal basis set). Mazziotti \cite{maz_prl} then developed an algorithm that transforms the SDP into a non-linear optimization program solved by a gradient-only method. This reduced the cost of the storage and the basic floating point operations, but at the cost of these nice convergence properties of the interior point methods.
In this paper we adapt a standard primal-dual interior point algorithm \cite{sturm} to the specific case of v2DM, in an attempt to retain the nice convergence properties, while reducing the storage and computational cost. In Sec.~\ref{v2DM} we present an introduction to the theory of $N$-representability, v2DM and some mathematical properties of the constraints. In Sec.~\ref{SDP} we discuss the representation of the problem as a primal-dual semidefinite program, and introduce the method we use to solve it. Then we apply the algorithm to a BCS (Bardeen-Cooper-Shrieffer) \cite{BCS} or pairing-type Hamiltonian in Sec.~\ref{app} and present the physical results and computational aspects. A summary is provided in Sec.~\ref{sum}.
\section{\label{v2DM} Variational density matrix determination}
When only two-body interactions are present, the Hamiltonian of a physical system can be written as:
\begin{equation}
\hat{H} = \sum_{\alpha\gamma}t_{\alpha\gamma} a^\dagger_\alpha a_\gamma + \frac{1}{4}\sum_{\alpha\beta\gamma\delta}V_{\alpha\beta;\gamma\delta}a^\dagger_\alpha a^\dagger_\beta a_\delta a_\gamma~,
\end{equation}
using second quantized notation where $a^\dagger_\alpha$ ($a_\alpha$) creates (annihilates) a fermion in a single-particle (sp) state $\alpha$ \cite{bijbel}.
The expectation value of the energy in an arbitrary $N$-particle state $\ket{\Psi^N}$ can be expressed in terms of the 2DM only,
\begin{equation}
E(\Gamma) = \mathrm{Tr}~\Gamma H^{(2)} = \sum_{\alpha<\beta;\gamma<\delta}\Gamma_{\alpha\beta;\gamma\delta}H^{(2)}_{\alpha\beta;\gamma\delta}~,
\label{ener_func}
\end{equation}
with the 2DM defined as:
\begin{equation}
\Gamma_{\alpha\beta;\gamma\delta} = \bra{\Psi^N}a^\dagger_\alpha a^\dagger_\beta a_\delta a_\gamma \ket{\Psi^N}~,
\label{2DM}
\end{equation}
and the reduced two-particle Hamiltonian,
\begin{equation}
H^{(2)}_{\alpha\beta;\gamma\delta} = \frac{1}{N-1}\left(\delta_{\alpha\gamma}t_{\beta\delta} - \delta_{\alpha\delta}t_{\beta\gamma} - \delta_{\beta\gamma}t_{\alpha\delta} + \delta_{\beta\delta}t_{\alpha\gamma}\right) + V_{\alpha\beta;\gamma\delta}~.
\end{equation}
The idea of v2DM is to determine the ground-state energy and other two- or one-body properties by minimizing the energy (\ref{ener_func}) using the 2DM as a variable. The 2DM is a much more compact object than the wave function because one keeps the dimension of two-particle (tp) space, no matter how many particles are involved. The problem is that there is no straightforward way to know whether an arbitrary matrix in tp-space $\Gamma$ is derivable from a physical wave function as in Eq.~(\ref{2DM}). Actually, it is sufficient that $\Gamma$ is derivable from an ensemble of $N$-particle wave functions, and this is called the $N$-representability problem \cite{coleman}. Some obvious necessary $N$-representability constraints are apparent from the definition (\ref{2DM}):
\begin{eqnarray}
\text{trace condition}\qquad\mathrm{Tr}~\Gamma &=& \sum_{\alpha<\beta}\Gamma_{\alpha\beta;\alpha\beta}=\frac{N(N-1)}{2}~,\\
\text{antisymmetry}\qquad\Gamma_{\alpha\beta;\gamma\delta} &=& -\Gamma_{\beta\alpha;\gamma\delta} = -\Gamma_{\alpha\beta;\delta\gamma} = \Gamma_{\beta\alpha;\delta\gamma}~,\\
\text{Hermiticity}\qquad\Gamma_{\alpha\beta;\gamma\delta} &=& \Gamma_{\gamma\delta;\alpha\beta}~,
\end{eqnarray}
but it turns out that there are many non-trivial constraints needed to ensure that a 2DM is physical.
\subsection{$N$-representability}
The necessary and sufficient conditions for $N$-representability are formally known \cite{payers}. A tp-matrix is $N$-representable if and only if, for every two-body Hamiltonian $\hat{H}_\nu$, the following inequality is satisfied:
\begin{equation}
\mathrm{Tr}~H^{(2)}_\nu \Gamma \geq E_0(H_\nu)~,
\end{equation}
where $E_0(H_\nu)$ is the exact $N$-particle ground-state energy corresponding to the Hamiltonian.
This is hardly a practical approach, as one needs to know the ground-state energy of every two-body Hamiltonian. Therefore one resorts to certain classes of Hamiltonians for which a lower bound to the ground-state energy is known. A Hamiltonian class that is used as necessary constraint is
\begin{equation}
\label{stand_constr_tp}
\bra{\Psi^N}B^\dagger B\ket{\Psi^N} \geq 0~,
\end{equation}
which leads to positivity conditions of linear matrix maps of the 2DM. If we want (\ref{stand_constr_tp}) to be restricted to tp-space there are three possible forms of the operator $B^\dagger$, leading to three conditions on the density matrix:
\paragraph{$B^\dagger = \sum_{\alpha\beta}p_{\alpha\beta}a^\dagger_\alpha a^\dagger_\beta$} leads to the trivial $\mathcal{P}$-condition:
\begin{equation}
\mathcal{P}(\Gamma) = \Gamma \succeq 0~,
\end{equation}
which imposes positive semidefiniteness on the 2DM.
\paragraph{$B^\dagger = \sum_{\alpha\beta}q_{\alpha\beta}a_\alpha a_\beta$} leads to the $\mathcal{Q}$-condition:
\begin{equation}
\mathcal{Q}(\Gamma) \succeq 0~,
\end{equation}
where the linear matrix map $\mathcal{Q}$ is defined as
\begin{eqnarray}
\nonumber\mathcal{Q}(\Gamma)_{\alpha\beta;\gamma\delta} &=& \bra{\Psi^N}a_\alpha a_\beta a^\dagger_\delta a^\dagger_\gamma\ket{\Psi^N}\\
\nonumber&=&\Gamma_{\alpha\beta;\gamma\delta} + (\delta_{\alpha\gamma}\delta_{\beta\delta} - \delta_{\alpha\delta}\delta_{\beta\delta})\frac{\bar{\bar{\Gamma}}}{N(N-1)}\\
&&- \delta_{\alpha\gamma}\rho_{\beta\delta} + \delta_{\alpha\delta}\rho_{\beta\gamma} + \delta_{\beta\gamma}\rho_{\alpha\delta} - \delta_{\beta\delta}\rho_{\alpha\gamma}~,
\label{Q}
\end{eqnarray}
with
\begin{equation}
\rho_{\alpha\gamma} = \frac{1}{N-1}\bar{\Gamma}_{\alpha\gamma} = \frac{1}{N-1}\sum_{\beta}\Gamma_{\alpha\beta;\gamma\beta}~,\\
\end{equation}
the one-body reduced density matrix (1DM), and with
\begin{equation}
\bar{\bar{\Gamma}} = \sum_{\alpha\beta}\Gamma_{\alpha\beta;\alpha\beta}~,
\end{equation}
the unrestricted trace of the 2DM.
\paragraph{$B^\dagger = \sum_{\alpha\beta}g_{\alpha\beta}a^\dagger_\alpha a_\beta$} which leads to the $\mathcal{G}$-condition:
\begin{equation}
\mathcal{G}(\Gamma) \succeq 0~,
\end{equation}
with the linear matrix map $\mathcal{G}$ defined as
\begin{eqnarray}
\nonumber\mathcal{G}(\Gamma)_{\alpha\beta;\gamma\delta} &=& \bra{\Psi^N}a^\dagger_\alpha a_\beta a^\dagger_\delta a_\gamma\ket{\Psi^N}\\
&=& \delta_{\beta\delta}\rho_{\alpha\gamma} - \Gamma_{\alpha\delta;\gamma\beta}~.
\label{G_up}
\end{eqnarray}
Another Hamiltonian class for which a lower bound to the ground-state energy is known gives rise to the so-called three-index conditions:
\begin{equation}
\bra{\Psi^N}\left\{B^\dagger,B\right\}\ket{\Psi^N} \geq 0~.
\label{three_index}
\end{equation}
In this article we will use two conditions that come from Eq.~(\ref{three_index}). 
\paragraph{$B^\dagger = \sum_{\alpha\beta\gamma}t_{\alpha\beta\gamma}a^\dagger_\alpha a^\dagger_\beta a^\dagger_\gamma$}
leads to the $\mathcal{T}_1$-condition:
\begin{equation}
\mathcal{T}_1(\Gamma) \succeq 0~,
\end{equation}
with the linear matrix map $\mathcal{T}_1$ defined as
\begin{eqnarray}
\nonumber\mathcal{T}_1\left(\Gamma\right)_{\alpha\beta\gamma;\delta\epsilon\zeta} &=& \bra{\Psi^N}a^\dagger_\alpha a^\dagger_\beta a^\dagger_\gamma a_\zeta a_\epsilon a_\delta + a_\alpha a_\beta a_\gamma a^\dagger_\zeta a^\dagger_\epsilon a^\dagger_\delta\ket{\Psi^N}\\
\nonumber&=&\left(\delta_{\gamma\zeta}\delta_{\beta\epsilon}\delta_{\alpha\delta} - \delta_{\gamma\epsilon}\delta_{\alpha\delta}\delta_{\beta\zeta} + \delta_{\alpha\zeta}\delta_{\gamma\epsilon}\delta_{\beta\delta} - \delta_{\gamma\zeta}\delta_{\alpha\epsilon}\delta_{\beta\delta} + \delta_{\beta\zeta}\delta_{\alpha\epsilon}\delta_{\gamma\delta} -\delta_{\alpha\zeta}\delta_{\beta\epsilon}\delta_{\gamma\delta}\right)\frac{\bar{\bar{\Gamma}}}{N(N - 1)}\\
\nonumber&& -\left(\delta_{\gamma\zeta}\delta_{\beta\epsilon} - \delta_{\beta\zeta}\delta_{\gamma\epsilon}\right)\rho_{\alpha\delta} + \left(\delta_{\gamma\zeta}\delta_{\alpha\epsilon} - \delta_{\alpha\zeta}\delta_{\gamma\epsilon}\right)\rho_{\beta\delta} - \left(\delta_{\beta\zeta}\delta_{\alpha\epsilon}- \delta_{\alpha\zeta}\delta_{\beta\epsilon}\right)\rho_{\gamma\delta}\\
\nonumber&& + \left(\delta_{\gamma\zeta}\delta_{\beta\delta} - \delta_{\beta\zeta}\delta_{\gamma\delta}\right)\rho_{\alpha\epsilon} - \left(\delta_{\gamma\zeta}\delta_{\alpha\delta} - \delta_{\alpha\zeta}\delta_{\gamma\delta}\right)\rho_{\epsilon\beta} + \left(\delta_{\beta\zeta}\delta_{\alpha\delta} - \delta_{\alpha\zeta}\delta_{\beta\delta}\right)\rho_{\gamma\epsilon}\\
\nonumber&& - \left(\delta_{\beta\delta}\delta_{\gamma\epsilon} - \delta_{\beta\epsilon}\delta_{\gamma\delta}\right)\rho_{\alpha\zeta} + \left(\delta_{\gamma\epsilon}\delta_{\alpha\delta} - \delta_{\alpha\epsilon}\delta_{\gamma\delta}\right)\rho_{\beta\zeta} - \left(\delta_{\beta\epsilon}\delta_{\alpha\delta} - \delta_{\alpha\epsilon}\delta_{\beta\delta}\right)\rho_{\gamma\zeta}\\
\nonumber&&+ \delta_{\gamma\zeta}\Gamma_{\alpha\beta;\delta\epsilon} - \delta_{\beta\zeta}\Gamma_{\alpha\gamma;\delta\epsilon} + \delta_{\alpha\zeta}\Gamma_{\beta\gamma;\delta\epsilon} - \delta_{\gamma\epsilon}\Gamma_{\alpha\beta;\delta\zeta} + \delta_{\beta\epsilon}\Gamma_{\alpha\gamma;\delta\zeta} - \delta_{\alpha\epsilon}\Gamma_{\beta\gamma;\delta\zeta}\\
&& + \delta_{\gamma\delta}\Gamma_{\alpha\beta;\epsilon\zeta} - \delta_{\beta\delta}\Gamma_{\alpha\gamma;\epsilon\zeta} + \delta_{\alpha\delta}\Gamma_{\beta\gamma;\epsilon\zeta}~.
\label{T1_up}
\end{eqnarray}
\paragraph{$B^\dagger = \sum_{\alpha\beta\gamma}t_{\alpha\beta\gamma}a^\dagger_\alpha a^\dagger_\beta a_\gamma$}
leads to the $\mathcal{T}_2$-condition
\begin{equation}
\mathcal{T}_2(\Gamma) \succeq 0~,
\end{equation}
with the linear matrix map $\mathcal{T}_2$ defined as
\begin{eqnarray}
\label{T2_up}\mathcal{T}_2(\Gamma)_{\alpha\beta\gamma;\delta\epsilon\zeta} &=& \bra{\Psi^N}a^\dagger_\alpha a^\dagger_\beta a_\gamma a^\dagger_\zeta a_\epsilon a_\delta + a^\dagger_\gamma a_\beta a_\alpha a^\dagger_\delta a^\dagger_\epsilon a_\zeta\ket{\Psi^N}\\
\nonumber&=& \left(\delta_{\alpha\delta}\delta_{\beta\epsilon} - \delta_{\alpha\epsilon}\delta_{\beta\delta}\right)\rho_{\gamma\zeta} + \delta_{\gamma\zeta}\Gamma_{\alpha\beta;\delta\epsilon} - \delta_{\alpha\delta}\Gamma_{\gamma\epsilon;\zeta\beta} + \delta_{\beta\delta}\Gamma_{\gamma\epsilon;\zeta\alpha} + \delta_{\alpha\epsilon}\Gamma_{\gamma\delta;\zeta\beta} - \delta_{\beta\epsilon}\Gamma_{\gamma\delta;\zeta\alpha}~.
\end{eqnarray}
The optimization problem that we have to solve can be summarized as:
\begin{equation}
\min_{\Gamma} \mathrm{Tr}~\Gamma H^{(2)}~,
\end{equation}
under the condition that
\begin{eqnarray}
\mathrm{Tr}~\Gamma &=& \frac{N(N-1)}{2}~,\\
\mathcal{L}(\Gamma) &\succeq& 0 \qquad \forall \mathcal{L} \in \{\mathcal{P,Q,G},\mathcal{T}_1,\mathcal{T}_2\}~.
\end{eqnarray}
\subsection{Hermitian adjoint maps}
For the following it is useful to introduce the Hermitian adjoints of matrix maps introduced in the previous section. The Hermitian adjoint maps are defined through:
\begin{equation}
\label{gen_herm}
\mathrm{Tr}~\mathcal{L}_i(\Gamma) A = \mathrm{Tr}~\mathcal{L}_i^\dagger(A)\Gamma~,
\end{equation}
in which $A$ is a matrix of the same dimension as the image of the map $\mathcal{L}_i$ in question (\emph{e.g.} a three-particle matrix for a $\mathcal{T}_1$ map, \emph{etc.}), and the traces sum over the appropriate indices.
The $\mathcal{P}$ and $\mathcal{Q}$ maps are Hermitian, so they are identical to their Hermitian adjoints. For the other maps however this is not the case. Using Eq.~(\ref{gen_herm}) the Hermitian adjoint of the $\mathcal{G}$ map can be shown to have the form:
\begin{eqnarray}
\label{G_down}
\mathcal{G}^\dagger\left(A\right)_{\alpha\beta;\gamma\delta} &=& \frac{1}{N-1}\left[\delta_{\beta\delta}\bar{A}_{\alpha\gamma} - \delta_{\alpha\delta}\bar{A}_{\beta\gamma} - \delta_{\beta\gamma}\bar{A}_{\alpha\delta} + \delta_{\alpha\gamma}\bar{A}_{\beta\delta}\right]\\
\nonumber&&\qquad\qquad - A_{\alpha\delta;\gamma\beta} + A_{\beta\delta;\gamma\alpha} + A_{\alpha\gamma;\delta\beta} - A_{\beta\gamma;\delta\alpha}~,
\end{eqnarray}
in which a particle-hole matrix $A$ is mapped onto tp-matrix space and
\begin{equation}
\bar{A}_{\alpha\gamma} = \sum_\lambda A_{\alpha\lambda;\gamma\lambda}~.
\end{equation}
The $\mathcal{T}_1$-operator maps a tp-matrix onto a three-particle matrix, so its Hermitian adjoint has to map a three-particle matrix $A$ onto tp-space. Solving Eq.~(\ref{gen_herm}) with $\mathcal{L} = \mathcal{T}_1$ one finds that:
\begin{eqnarray}
\label{T1_down}
\mathcal{T}^\dagger_1\left(A\right)_{\alpha\beta;\gamma\delta} &=& \frac{2}{N(N-1)}\left(\delta_{\alpha\gamma}\delta_{\beta\delta} - \delta_{\alpha\delta}\delta_{\beta\gamma}\right)\mathrm{Tr}~A + \bar{A}_{\alpha\beta;\gamma\delta}\\
\nonumber&&-\frac{1}{2(N-1)}\left[\delta_{\beta\delta}\bar{\bar{A}}_{\alpha\gamma} - \delta_{\alpha\delta}\bar{\bar{A}}_{\beta\gamma} - \delta_{\beta\gamma}\bar{\bar{A}}_{\alpha\delta} + \delta_{\alpha\gamma}\bar{\bar{A}}_{\beta\delta}\right]~,
\end{eqnarray}
with
\begin{eqnarray}
\bar{A}_{\alpha\beta;\gamma\delta} &=& \sum_\lambda A_{\alpha\beta\lambda;\gamma\delta\lambda}~,\\
\bar{\bar{A}}_{\alpha\gamma} &=& \sum_{\lambda\kappa} A_{\alpha\lambda\kappa;\gamma\lambda\kappa}~.
\end{eqnarray}
In the same way one can derive for $\mathcal{L}=\mathcal{T}_2$ that
\begin{eqnarray}
\label{T2_down}
\mathcal{T}^\dagger_2(A)_{\alpha\beta;\gamma\delta} &=& \frac{1}{2(N-1)}\left[\delta_{\beta\delta}\tilde{\tilde{A}}_{\alpha\gamma} - \delta_{\alpha\delta}\tilde{\tilde{A}}_{\beta\gamma} - \delta_{\beta\gamma}\tilde{\tilde{A}}_{\alpha\delta} + \delta_{\alpha\gamma}\tilde{\tilde{A}}_{\beta\delta}\right] + \bar{A}_{\alpha\beta;\gamma\delta}\\
\nonumber&&-\left[\tilde{A}_{\delta\alpha;\beta\gamma} - \tilde{A}_{\delta\beta;\alpha\gamma} - \tilde{A}_{\gamma\alpha;\beta\delta} + \tilde{A}_{\gamma\beta;\alpha\delta}\right]~,
\end{eqnarray}
 with this time $A$ a matrix on two-particle-one-hole space and
\begin{eqnarray}
\tilde{\tilde{A}}_{\alpha\gamma} &=& \sum_{\lambda\kappa}A_{\lambda\kappa\alpha;\lambda\kappa\gamma}~,\\
\bar{A}_{\alpha\beta;\gamma\delta} &=& \sum_{\lambda}A_{\alpha\beta\lambda;\gamma\delta\lambda}~,\\
\tilde{A}_{\alpha\beta;\gamma\delta} &=& \sum_{\lambda}A_{\lambda\alpha\beta;\lambda\gamma\delta}~.
\end{eqnarray}
\section{\label{SDP} Primal-dual semidefinite program}
The variational method described in the previous section can be formulated as a primal-dual semidefinite program. A general 2DM, describing an $N$-particle system can be expanded in an arbitrary orthogonal basis $\{f^i\}$ of traceless matrix space as
\begin{equation}
\Gamma = \frac{N(N - 1)}{M(M-1)} \mathbb{1}_{\text{tp}} + \sum_i \gamma_i f^i~,
\end{equation}
with $M$ the dimension of single-particle (sp) space, and the unit matrix on tp space defined as
\begin{equation}
\left(\mathbb{1}_{\text{tp}}\right)_{\alpha\beta;\gamma\delta} = \delta_{\alpha\gamma}\delta_{\beta\delta} - \delta_{\alpha\delta}\delta_{\beta\gamma}~.
\end{equation}
The energy of the system can be written as a function of the $\gamma$'s as
\begin{equation}
\mathrm{Tr}~\Gamma H^{(2)} = \frac{N(N - 1)}{M(M-1)}\mathrm{Tr}~H^{(2)} + \sum_i \gamma_i \mathrm{Tr}~H^{(2)}f^i~.
\end{equation}
Because the necessary $N$-representability conditions can be written as linear homogeneous matrix maps of $\Gamma$, we can also write them as a function of the $\gamma$'s:
\begin{equation}
\mathcal{L}\left({\Gamma}\right) = \frac{N(N - 1)}{M(M-1)}\mathcal{L}\left({\mathbb{1}_\text{tp}}\right) + \sum_{i} \gamma_i \mathcal{L}\left({f^i}\right) \succeq 0~.
\end{equation}
If we now consider the direct sum of the linear spaces associated with the maps and define the block matrices:
\begin{equation}
u^0 = \frac{N(N - 1)}{M(M-1)}\bigoplus_k \mathcal{L}_k\left(\mathbb{1}_\text{tp}\right) \qquad\text{and}\qquad u^i = \bigoplus_k \mathcal{L}_k\left(f^i\right)~,
\end{equation}
then we can formulate v2DM as a standard dual semidefinite program \cite{vandenberghe}:
\begin{equation}
\min_\gamma~\gamma^T h \qquad \text{on condition that} \qquad Z = u^0 + \sum_i \gamma_i u^i \succeq 0~,
\label{primal}
\end{equation}
in which $h^i = \mathrm{Tr}~H^{(2)}f^i$. The primal problem corresponding to (\ref{primal}) optimizes the matrixvariable $X$, the problem being defined as:
\begin{equation}
\max_X~ \left(-\mathrm{Tr}~Xu^0\right) \qquad \text{on condition that} \qquad \mathrm{Tr}~Xu^i = h^i \qquad \text{and} \qquad X\succeq 0~.
\label{dual}
\end{equation}
$X$ will be a block matrix because the $u$-matrices are block matrices. The primal-dual gap $\eta$ is defined as the difference between the primal and the dual cost function for a certain primal-dual point $(X,Z)$:
\begin{equation}
\eta = \gamma^T h + \mathrm{Tr}~u^0 X = \sum_i \gamma_i \mathrm{Tr}~X u^i + \mathrm{Tr}~X u^0 = \mathrm{Tr}~X Z \geq 0~,
\end{equation}
as $X$ and $Z$ are positive semidefinite matrices. We can see that the smallest value of $\eta$ will be reached when both the primal and the dual problem are optimal. It can be proven that if the primal and the dual problem are both strictly feasible, then the primal-dual gap vanishes at their solution \cite{vandenberghe}. This means that the primal-dual gap can be used as a convergence criterion for the algorithm. Even better, at any point during the optimization, the error on the current value is limited from above by the primal-dual gap. Note that in our previous implementation \cite{atomic}, a dual-only algorithm was used. The properties of the present primal-dual method can lead to a serious reduction in computation time since we can stop the algorithm at a prescribed error estimate.
\subsection{Equations of motion}
There are several known methods to solve a semidefinite program. In this paper a path-following interior point method is used. The central path is defined as the set of primal-dual points for which
\begin{equation}
\label{cent_path}
X Z = \frac{\eta}{n} \mathbb{1}_\text{sup}~,
\end{equation}
with $n$ the total dimension of the $X$ and $Z$ matrices and $\mathbb{1}_\text{sup}$ the direct sum of the unity matrices on the different constraint spaces:
\begin{equation}
\mathbb{1}_\text{sup} = \bigoplus_k \mathbb{1}_k~.
\end{equation}
In the path-following algorithm \cite{sturm} we try to follow the central path, reducing the primal-dual gap along the way. Consider a primal-dual point $(X,Z)$ on the central path with primal-dual gap $\eta$. We want to know what is the primal-dual point on the central path with primal-dual gap scaled down with a factor $\nu$. Rephrasing, we are looking for the $(\Delta_X,\Delta_Z)$ that solve:
\begin{equation}
(X + \Delta_X)(Z + \Delta_Z) = \frac{\nu\eta}{n}\mathbb{1}_\text{sup}~.
\label{EOM}
\end{equation}
There are several ways to symmetrize these equations. Using the method proposed by \cite{sturm}, two equivalent equations (called the dual and the primal) are obtained, \emph{i.e.} one has to solve the equations
\begin{eqnarray}
\label{P_eom}(\text{dual}) : \Delta_X + D^{-1}\Delta_Z D^{-1} &=& \frac{\nu\eta}{n} Z^{-1} - X~,\\
\label{D_eom}(\text{primal}) : \Delta_Z + D~\Delta_{X}~D &=& \frac{\nu\eta}{n} X^{-1} - Z~,
\end{eqnarray}
and under the condition that:
\begin{equation}
\label{eom_constr}
\mathrm{Tr}~\Delta_Xu^i = 0 \qquad \text{and} \qquad \Delta_Z = \sum_i \left(\delta \gamma\right)_i u^i~,
\end{equation}
and with 
\begin{equation}
D(X,Z) = X^{-\frac{1}{2}}\left(X^{\frac{1}{2}}ZX^{\frac{1}{2}}\right)^{\frac{1}{2}}X^{-\frac{1}{2}}~.
\label{metric}
\end{equation}
\subsubsection{Solution to the dual equation}
In order to obtain the primal-dual direction $(\Delta_X,\Delta_Z)$ , the dual equation (\ref{P_eom}) is first projected onto the space spanned by the non-orthogonal basis $\{u^i\}$ (which we will call $\mathcal{U}$-space). With $B$ denoting the right-hand side of (\ref{P_eom}) and making use of Eq.~(\ref{eom_constr}) we obtain:
\begin{equation}
\label{primal_cg}
\sum_j \underbrace{\left(\mathrm{Tr}~D^{-1} u^j D^{-1} u^i\right)}_{\mathcal{H}^D_{ij}} \Delta\gamma_j = \mathrm{Tr}~B u^i~,
\end{equation}
which can be seen to be a symmetrical, positive-definite linear system and as such can be solved iteratively using the linear conjugate gradient method. This can be done without explicit construction of the dual Hessian matrix $\mathcal{H}^D$ or any reference to the non-orthogonal basis set $\{u^i\}$. This is because $\mathcal{H}^D$ can be seen as a map from traceless tp-matrix space onto itself, by using the Hermitian adjoints of the linear maps $\mathcal{L}$. Consider an arbitrary traceless tp-matrix:
\begin{equation}
\epsilon = \sum_j \epsilon_j f^j~.
\end{equation}
Using (\ref{gen_herm}) and the fact that the $\mathcal{L}$'s are linear and homogeneous we obtain that the image of $\epsilon$ under the dual Hessian map can be written as:
\begin{equation}
\mathcal{H}^D\epsilon = \hat{P}_{\text{Tr}}\left[\sum_k\mathcal{L}^\dagger_k\left(D_k^{-1}\mathcal{L}_k\left(\epsilon\right) D_k^{-1}\right)\right]~,
\end{equation}
in which the $D_k$ are the blocks of the $D$ matrix corresponding to the different constraints $\mathcal{L}_k$, and $\hat{P}_\text{Tr}$ stands for the projection operator onto traceless tp-matrix space:
\begin{equation}
\hat{P}_{\text{Tr}}(A) = A - \frac{2\mathrm{Tr}~A}{M(M-1)}\mathbb{1}_\text{tp}~.
\end{equation}
\subsubsection{Solution to the primal equation}
The solution of the primal equation (\ref{D_eom}) is obtained in the same manner, by projecting this equation onto $\mathcal{C}$-space, the orthogonal complement of $\mathcal{U}$-space. With $B$ denoting the right-hand side of the equation (\ref{D_eom}) and making use of Eq.~(\ref{eom_constr}) one gets:
\begin{equation}
\label{dual_cg}
\sum_j \underbrace{\left(\mathrm{Tr}~D~c^j~D~c^i\right)}_{\mathcal{H}^P_{ij}} \delta x_j = \mathrm{Tr}~B c^i~,
\end{equation}
where we have used
\begin{equation}
\Delta_X = \sum_i \delta x_i~c^i~.
\end{equation}
This is again a symmetrical positive-definite system of linear equations that can be solved iteratively using the linear conjugate gradient method. As with the dual equation it can be solved without explicit construction of the Hessian matrix $\mathcal{H}^P$, or any reference to the basisset $\{c^i\}$, because $\mathcal{H}^P$ can be seen as a map from $\mathcal{C}$-space onto itself. For an arbitrary matrix in $\mathcal{C}$-space:
\begin{equation}
\epsilon = \sum_i \epsilon_i c^i~,
\end{equation}
the image of $\epsilon$ under the primal Hessian map is
\begin{equation}
\mathcal{H}^P\epsilon = \hat{P}_{\mathcal{C}}\left[D\epsilon D\right]~.
\end{equation}
in which $\hat{P}_\mathcal{C}$ is the projection onto $\mathcal{C}$-space. This projection can be executed quickly by using the inverse of the overlap matrix of the $\mathcal{U}$-space basis vectors. Suppose we have an arbitrary block matrix $A$ of the same dimension as $X$ and $Z$. First we project it onto the space spanned by the basis $\{u^0,u^i\} = \{u^\alpha\}$. The projected matrix $A'$ reads as:
\begin{equation}
A' = \sum_{\alpha\beta} \mathrm{Tr}~\left[Au^\alpha\right] \left(\mathcal{S}^{-1}\right)_{\alpha\beta}u^\beta~,
\end{equation}
where the overlap matrix $\mathcal{S}$ appears because of the non-orthogonality of the basis.
Due to the special properties of the linear matrix maps $\mathcal{L}$ that determine the basis matrices $u^\alpha$, the inverse overlap matrix can also be considered as a map from tp space onto itself (see Appendix \ref{overlapmatrix} and \ref{inverse_overlapmatrix} for the actual analytic expression of this map). The projected matrix $A'$ can now be written in block-matrix form as:
\begin{equation}
A' = \bigoplus_l \mathcal{L}_l\left[\mathcal{S}^{-1}\left(\sum_k \mathcal{L}^\dagger_k\left(A_k\right)\right)\right]~.
\end{equation}
To project $A$ onto $\mathcal{U}$-space we still have to remove the component along the $u^0$-matrix:
\begin{equation}
\hat{P}_{\mathcal{U}}A = A' - \left(\frac{\mathrm{Tr}~u^0 A'}{\mathrm{Tr}~u^0u^0}\right)u^0~.
\end{equation}
Since $\mathcal{C}$-space is the orthogonal complement of the $\mathcal{U}$-space, the desired projection of $A$ onto the $\mathcal{C}$-space is simply given by
\begin{equation}
\hat{P}_{\mathcal{C}}A = A - \hat{P}_{\mathcal{U}}A~.
\end{equation}
\subsection{Outline of the algorithm}
In this section a short outline of the algorithm will be presented. The first step is to initialize the primal-dual variables, after which they are directed towards the central path. Then the actual minimization of the primal-dual gap takes place, which is done in a predictor-corrector loop.
\subsubsection{Initialization}
We need a feasible primal-dual starting point. An initial feasible dual point $Z^{(0)}$, \emph{i.e.} a matrix that satisfies the inequality (\ref{primal}), is easily found by setting
\begin{equation}
Z^{(0)} = u^0~,
\end{equation}
which corresponds to setting al the $\gamma_i$'s equal to zero. A feasible primal starting point will have to satisfy Eq.~(\ref{dual}). To construct such a point we take a completely random matrix $X$ and project it onto a matrix $X'$ for which
\begin{equation}
\label{projham}
\mathrm{Tr}~X'u^i = h^i~.
\end{equation}
This is again achieved using the inverse overlap matrix of the $\{u^\alpha\}$ basis,
\begin{equation}
X' = X - \underbrace{\sum_{\alpha\beta}\left(\mathrm{Tr}~Xu^\alpha - h^\alpha\right)\mathcal{S}^{-1}_{\alpha\beta}u^\beta}_{X^\perp}~.
\end{equation}
The last term on the right-hand side can be computed as:
\begin{equation}
X^\perp = \bigoplus_l \mathcal{L}_l\left[\mathcal{S}^{-1}\left(\sum_k \mathcal{L}^\dagger_k\left(X_k\right) - H^{(2)}\right)\right]~.
\end{equation}
At this point, $X'$ satifies the equality constraint (\ref{projham}), and one just has to add $u^{0}$, with a positive scaling factor that is large enough to ensure positive semidefiniteness:
\begin{equation}
X^{(0)} = X' + \alpha u^{0}\succeq 0~.
\end{equation}
\subsubsection{\label{centering}Centering run}
Before the actual program can be started, a couple of centering steps have to be taken, which is done by solving the equations (\ref{P_eom}) and (\ref{D_eom}) with $\nu = 1$. The purpose is to go sufficiently near the central path, without bothering about the primal-dual gap. In a first step, Eq.~(\ref{primal_cg}) which has the smallest dimension, is solved using the conjugate gradient method, and the dual solution $\Delta_Z$ is obtained. The primal solution $\Delta_X$ then follows from the dual equation (\ref{P_eom}) by substitution. For these initial centering steps, both linear systems are so well conditioned that hardly any iterations are needed for convergence. As a measure for the distance from the center we use the potential \cite{vandenberghe}:
\begin{equation}
\Phi(X,Z) = -\ln\det X - \ln \det Z~,
\end{equation}
which is minimal (for points with the same primal-dual gap $\eta = \mathrm{Tr}~XZ$) on the central path for which Eq.~(\ref{cent_path}) is satisfied:
\begin{equation}
\Phi(X^c,Z^c) = -n \ln{\frac{\eta}{n}}~.
\end{equation}
When the potential difference (which is always positive):
\begin{eqnarray}
\nonumber\Psi(X,Z) &=& \Phi(X,Z) - \Phi(X^c,Z^c)\\
\label{logpot}&=& {n}\ln \mathrm{Tr}~XZ -{n}\ln{n}-\ln\det X - \ln \det Z~,
\end{eqnarray}
is sufficiently small, the centering run is stopped.
\subsubsection{Predictor-corrector run}
In this part of the program the primal-dual gap is minimized by alternating predictor and corrector steps. A predictor step tries to reduce the primal-dual gap by solving the equations (\ref{P_eom}) and (\ref{D_eom}) with $\nu = 0$. This is done in exactly the same way as for the centering run, by first solving (\ref{primal_cg}) for $\Delta_Z$, then substituting into (\ref{P_eom}) to obtain an approximate primal step $\Delta_X$. The final primal step $\Delta_X$ is obtained by solving (\ref{dual_cg}) using the conjugate gradient method with the approximate $\Delta_X$ as a starting point. Note that when the primal-dual gap decreases, the condition number of the primal and dual Hessian matrices increases and more iterations are needed before convergence is reached. One can adjust the convergence criteria of the primal and dual conjugate gradient loops, in order to minimize the combined number of iterations.

At this point we have a predictor direction $(\Delta_X,\Delta_Z)$. The logarithmic potential $\phi(\alpha) = \Psi(X + \alpha \Delta_X,Z + \alpha \Delta_Z)$ in the predictor direction (see Eq.~(\ref{logpot})) can be simply evaluated for any value of $\alpha$ by precomputing the eigenvalues $\lambda^X_i$ of $X^{-\frac{1}{2}}\Delta_X X^{-\frac{1}{2}}$ and $\lambda^Z_i$ of $Z^{-\frac{1}{2}}\Delta_Z Z^{-\frac{1}{2}}$. One then has
\begin{eqnarray}
\phi(\alpha) &=& \Psi(X,Z) +  \ln\left[1 + \alpha (c_X + c_Z)\right] - \sum_i \ln (1 + \alpha \lambda^X_i) - \sum_i \ln(1 + \alpha \lambda^Z_i)~,
\end{eqnarray}
where
\begin{equation}
c_Z =  \frac{1}{\eta}\mathrm{Tr}~X\Delta_Z \qquad\text{and}\qquad c_X =  \frac{1}{\eta}\mathrm{Tr}~Z\Delta_X~,
\end{equation}
With a standard bisection method one can now compute the stepsize $\alpha$ corresponding to the maximal deviation from the central path we want to allow.

After the predictor step, a corrector step is taken, which is equivalent to the centering step described previously (see Sec. \ref{centering}). The alternation of predictor and corrector steps continues until the primal-dual gap is smaller then the desired value.
\begin{center}
\begin{figure}
\includegraphics[scale=0.6]{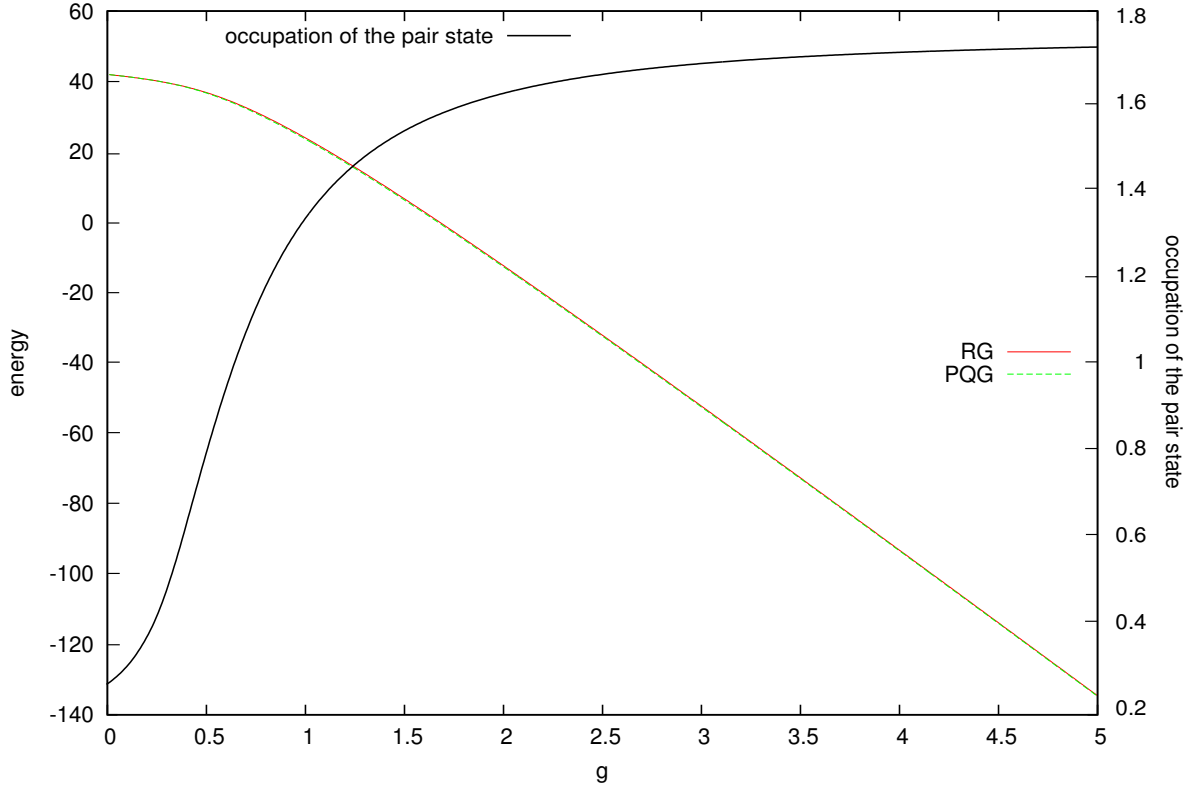}
\caption{\label{energy} The ground-state energy as calculated by v2DM(PQG) and the Richardson-Gaudin equations (RG), together with the pair occupation in the groundstate by v2DM(PQG), as a function of the pairing interaction strength $g$.}
\end{figure}
\end{center}
\begin{center}
\begin{figure}
\includegraphics[scale=0.6]{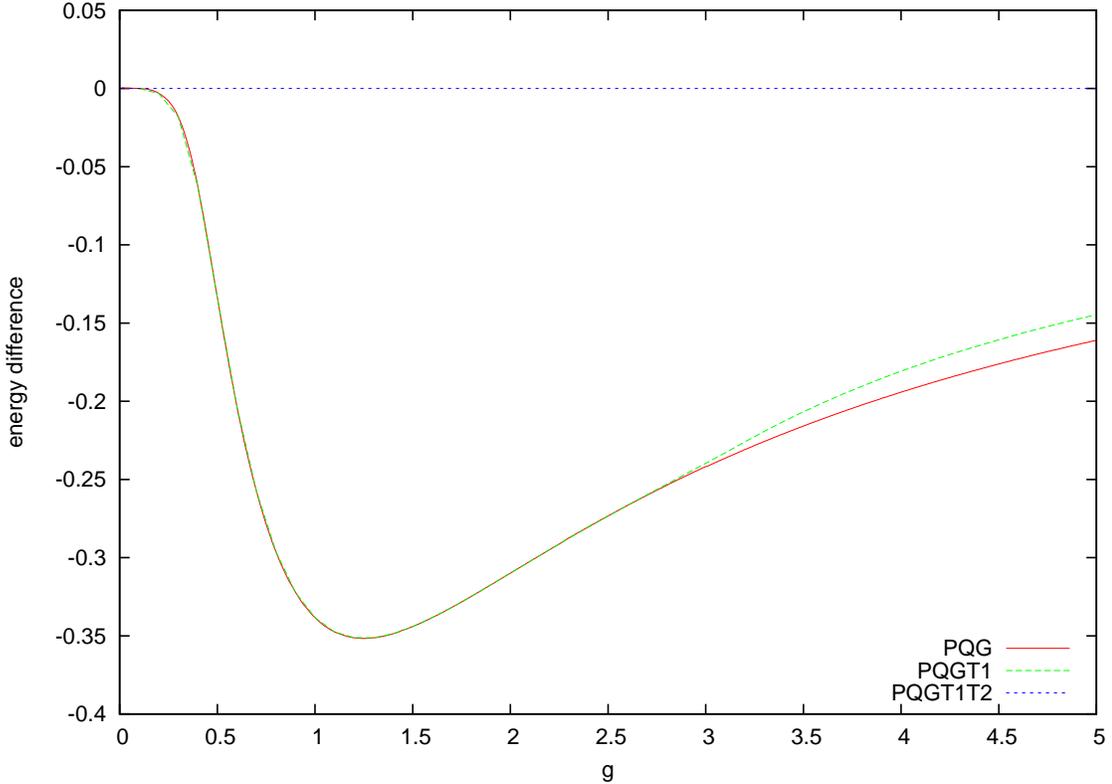}
\caption{\label{diff} The difference between the ground-state energy calculated by v2DM with various constraints, and the exact solution, as a function of pairing strength $g$.}
\end{figure}
\end{center}
\section{\label{app} Application to the BCS Hamiltonian}
\subsection{\label{bcsham}The BCS Hamiltonian}
The algorithm introduced in Sec. \ref{SDP} is applied to the BCS Hamiltionian \cite{BCS}. The BCS Hamiltonian is an interesting system that models the competition between a single-particle operator and a schematic pairing interaction:
\begin{equation}
\label{BCS_ham}
\hat{H} = \sum_{i\sigma}\epsilon_i a^\dagger_{i\sigma}a_{i\sigma} - g \sum_{ij} a^\dagger_{i\uparrow}a^\dagger_{i\downarrow}a_{j\downarrow}a_{j\uparrow}~.
\end{equation}
Here the single-particle levels are denoted with an index $i=1,\ldots,M$, and the up (down) spin as $\sigma = \uparrow(\downarrow)$.  When the pairing strength $g$ is small compared to the single-particle level spacing, the energy is minimized by filling up the single-particle orbitals up to the fermi level. With increasing $g$ however, it becomes advantageous to form pairs, \emph{i.e.} it is energetically favorable to maximize the ground-state occupation of the fermion pair state $\sum_i a^\dagger_{i\uparrow}a^\dagger_{i\downarrow}$. This problem is hard to solve using standard perturbative methods as these tend to break down when pairs are formed. An exact solution based on the Bethe-ansatz exists for this problem, however, and involves solving a system of non-linear equations \cite{richardson}. These equations are notoriously difficult to solve because, for certain critical values of $g$, the equations become singular. Several approaches have been suggested for solving these equations \cite{richardson2,rombouts}. In this paper we follow the approach recently proposed by De Baerdemacker \cite{stijn}. The exact ground-state energies as a function of $g$ are compared to the v2DM results calculated within the present formalism.
\subsection{Results}
We have studied the Hamiltonian Eq. (\ref{BCS_ham}) with $M=12$ doubly degenerate equidistant single-particle levels and $N=12$ fermions, and $g$ ranging from $0$ to $5$ in steps of $0.01$. v2DM calculations were performed with respectively $PQG$, $PQGT_1$ and $PQGT_1T_2$ constraints.
The resulting ground-state energy is compared to the exact solution in Fig.~\ref{energy}. For all values of $g$ the agreement is already remarkably good at the $PQG$ level. To appreciate how the result improves when constraints are added the difference between the various v2DM results and the exact solution is plotted in Fig.~\ref{diff}. Note that the difference is always negative, since v2DM provides a variational lower bound. As one observes, all approximations describe exactly the non-interacting small-$g$ limit. When $g$ becomes larger, there is competition between different types of ground states and the performance of $PQG$ gets worse up to $g\approx1.4$. For larger $g$ the $PQG$ result becomes better again. In fact, we checked (by omitting the single-particle piece) that also the $g\rightarrow\infty$ limit becomes exact for $PQG$, which is a peculiarity of the schematic pairing force. The $PQGT_1$ results show that the $T_1$ constraint only becomes active around $g = 2.5$, and ensures faster convergence to the exact $g\rightarrow\infty$ limit. Somewhat surprisingly, adding the $T_2$ condition is sufficient for obtaining the exact solution at all values of $g$.
\subsection{Computational Performance}
\begin{center}
\begin{figure}
\includegraphics[scale=0.6]{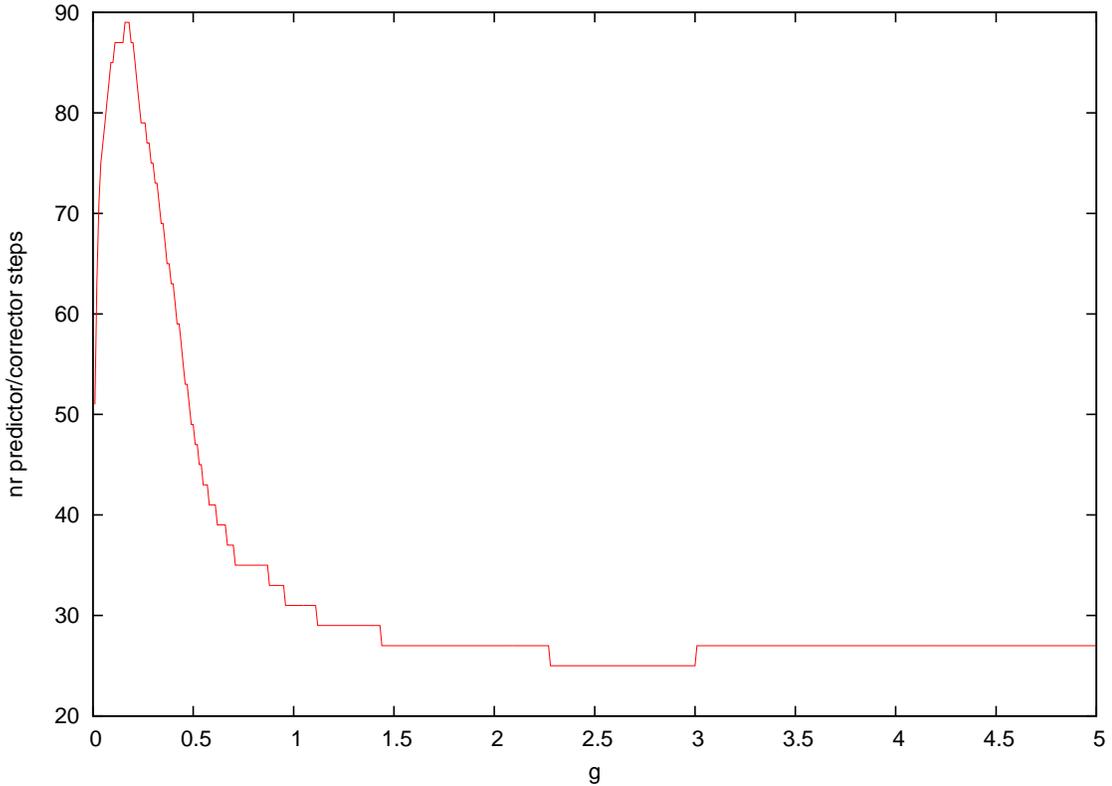}
\caption{\label{nr_newton_it} Number of predictor and corrector steps needed for convergence, as a function of the pairing strength $g$ in v2DM(PQG).}
\end{figure}
\end{center}

\begin{center}
\begin{figure}
\includegraphics[scale=0.6]{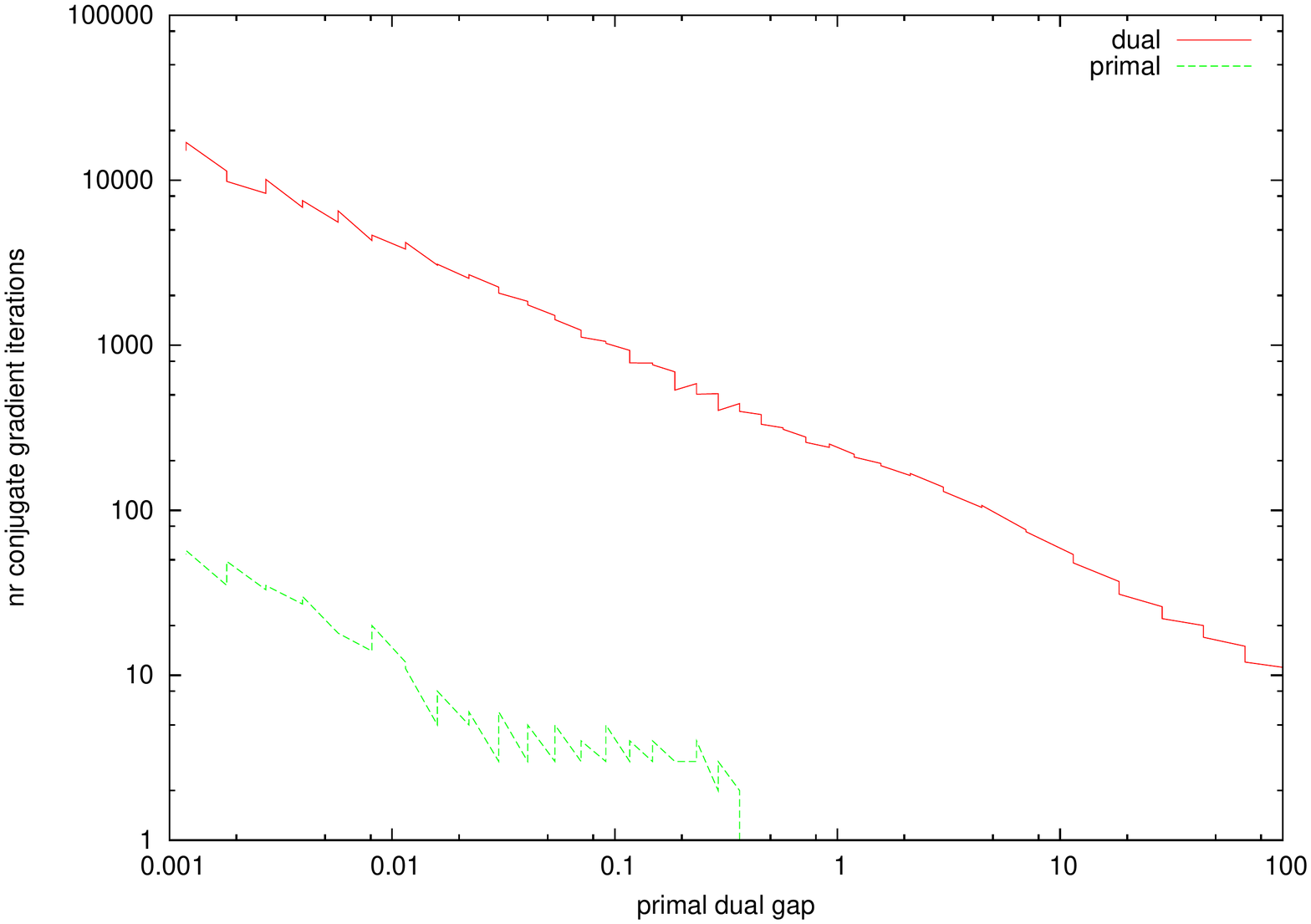}
\caption{\label{nr_iter_bad} Number of primal and dual conjugate gradient iterations needed for convergence, as a function of the primal-dual gap $\eta$ for $g=0.25$ in v2DM(PQG).}
\end{figure}
\end{center}

\begin{center}
\begin{figure}
\includegraphics[scale=0.6]{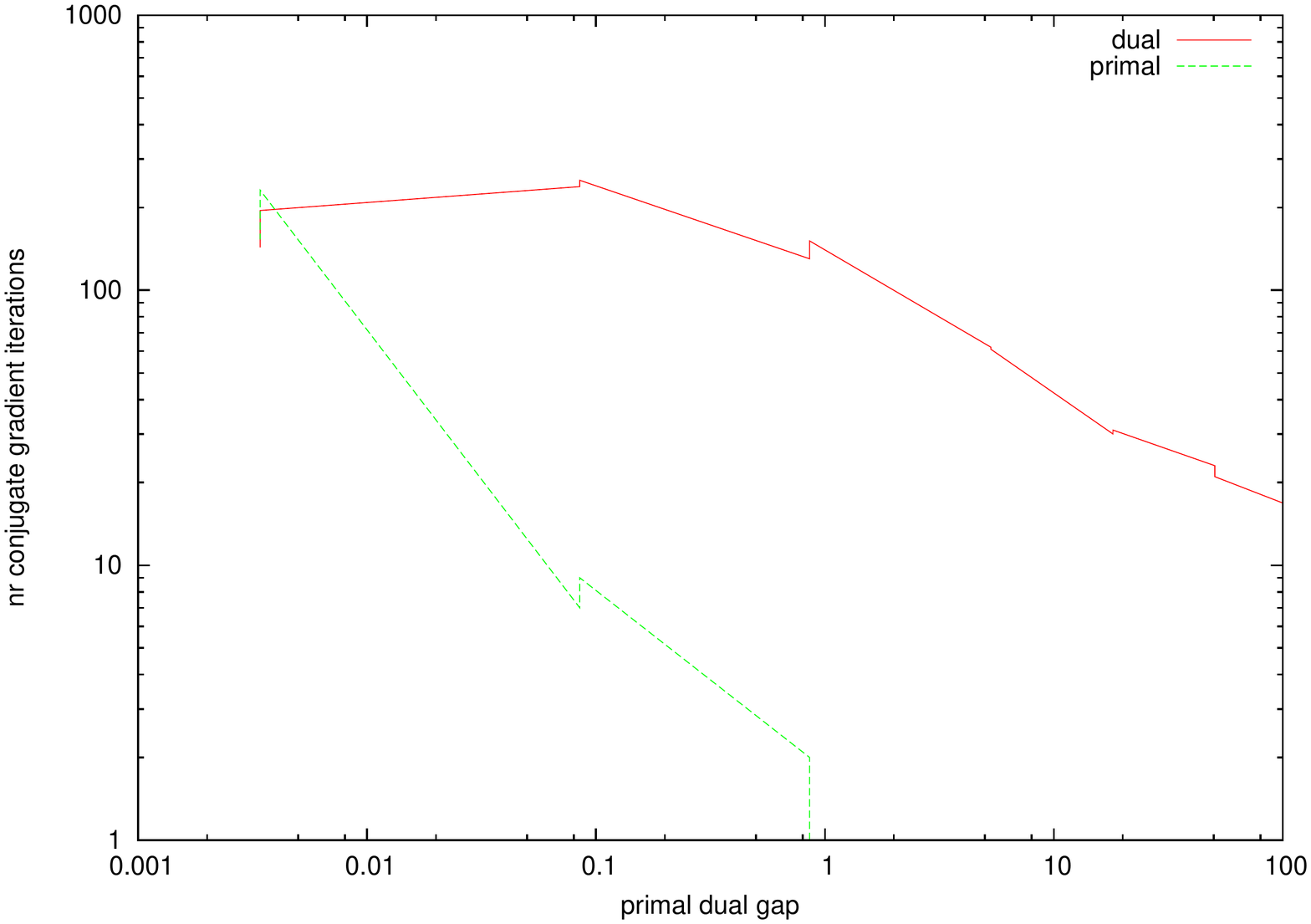}
\caption{\label{nr_iter_good} Number of primal and dual conjugate gradient iterations needed for convergence, as a function of the primal-dual gap $\eta$ for $g=4$ in v2DM(PQG).}
\end{figure}
\end{center}

Some of the computational aspects of the algorithm are worth pointing out. It is interesting to see \emph{e.g.} that depending on the pairing interaction parameter $g$, the convergence properties of the algorithm change. In Fig.~\ref{nr_newton_it} the joint number of predictor and corrector steps needed for convergence, is plotted as a function of $g$. One observes a sharp peak at fairly small $g$, just when the perturbative regime is left and the structure of the ground state changes. For $g=0.25$, which is at the position of the peak in Fig. \ref{nr_newton_it}, we have plotted in Fig. \ref{nr_iter_bad} the number of conjugate gradient iterations needed for convergence, of both the dual and the primal linear system, as a function of the primal-dual gap $\eta$. As expected, the number of iterations for the dual problem increases with decreasing primal-dual gap, as the linear system grows ill-conditioned. The primal conjugate gradient loop only becomes active for small values of $\eta$. This signals that the numerical stability becomes too small to generate a high quality approximation for $\Delta_X$ using the $\Delta_Z$ obtained in the dual conjugate gradient loop. Anyway, the needed number of primal iteration remains insignificant compared to the dual ones, for all values of $\eta$. The situation at $g=0.25$ is the worst case. For larger values of $g$, where the number of predictor-corrector steps is smaller and approximately constant (see Fig. \ref{nr_newton_it}), the number of conjugate gradient iterations is also drastically reduced. A typical behaviour is plotted in Fig. \ref{nr_iter_good} for $g=4$.
\section{\label{sum}Summary and discussion}
Interacting quantum many-particle systems lie at the heart of most issues in condensed matter, molecular/atomic and nuclear physics. Their analysis may be rephrased as the problem of minimizing the energy, expressed as a linear function of a two-body density matrix, subject to the $N$-representability constraint that the 2DM can be derived from a physical $N$-particle system. By working solely with the 2DM, rather than with the $N$-particle wave function itself, the problem of the exponentially exploding dimension of $N$-particle Hilbert space with increasing $N$ is circumvented. The complexity of the problem is shifted, however, to the characterization of the $N$-particle representable 2DM's. In practice, a limited set of necessary but not sufficient conditions for $N$-representability are imposed during the minimization, resulting in a strict lower bound to the energy, which converges to the exact energy when more and more $N$-representability conditions are imposed.

Commonly used $N$-representability conditions impose the positive semidefiniteness of a set of linear matrix functionals of the 2DM. In this way the quantum many-body problem is converted into a well established field of optimization techniques called semidefinite programming. Standard packages for SDP, however, fail to take into account properties of the physical problem that can be exploited.

Using specific mathematical properties of the constraints for the v2DM problem, we have adapted a standard primal-dual interior point method to be computationally cheaper, both in storage as in floating point operations. We make extensize use of the algebra of linear matrix maps to calculate efficiently some intermediate quantities. During the Newton minimization procedure, a new direction in 2DM space is found iteratively using the conjugate gradient algorithm, thereby exploiting the fact that the product of the Hessian with a 2DM is considerably cheaper for the physical problem at hand than in a general situation.

As an example we have applied the algorithm to a BCS-type Hamiltonian. We found that the standard constraints work very well for this kind of problem. The computational performance of the method was analyzed, and it was shown that the convergence behaviour is dependent on the value of the pairing strength parameter. As in our previous algorithm \cite{atomic} the method slows down near the solution, because the matrices involved become ill conditioned. The present primal-dual algorithm allows to control this since the primal-dual gap provides an upper bound to the remaining error. Therefore the algorithm can be stopped when the required accuracy is reached, saving many unnecessary iterations.
\section{Acknowledgements}
We gratefully acknowledge financial support from FWO-Flanders and the research council of Ghent University. We would like to thank Paul W. Ayers for his useful suggestions. B.V., H.V.A., P.B. and D.V.N. are Members of the QCMM alliance Ghent-Brussels.
\appendix
\section{\label{overlapmatrix} Calculation of the overlap-matrix map}
The overlap matrix of the non-orthogonal basisset $\{u^\alpha\}$ is defined as:
\begin{equation}
\mathcal{S}_{\alpha\beta} = \mathrm{Tr}~u^\alpha u^\beta~.
\end{equation}
Using the Hermitian adjoints of the linear maps $\mathcal{L}$ we can rewrite this as:
\begin{equation}
\mathcal{S}_{\alpha\beta} = \sum_k \mathrm{Tr}~\left[\mathcal{L}_k^\dagger\left(\mathcal{L}_k\left(f^\alpha\right)\right) f^\beta\right]~,
\end{equation}
in which $\{f^\alpha\}$ \emph{is} an orthogonal basis of tp-matrix space. This means that the overlap matrix can be seen as a linear map from tp-space onto itself, whose action onto a tp-matrix $\Gamma$ is:
\begin{equation}
\mathcal{S}\left(\Gamma\right) = \sum_k \mathcal{L}_k^\dagger\left(\mathcal{L}_k\left(\Gamma\right)\right)~.
\end{equation}
It turns out that this map can be written as a generalized $\mathcal{Q}$ map, which is defined as:
\begin{equation}
\mathcal{Q}(a,b,c)\left(\Gamma\right)_{\alpha\beta;\gamma\delta} = a\Gamma_{\alpha\beta;\gamma\delta} + b\left(\delta_{\alpha\gamma}\delta_{\beta\delta} - \delta_{\alpha\delta}\delta_{\beta\gamma}\right)\bar{\bar{\Gamma}} - c\left(\delta_{\alpha\gamma}\bar{\Gamma}_{\beta\delta} - \delta_{\beta\gamma}\bar{\Gamma}_{\alpha\delta} - \delta_{\alpha\delta}\bar{\Gamma}_{\beta\gamma} + \delta_{\beta\delta}\bar{\Gamma}_{\alpha\gamma}\right)~.
\label{Q_like}
\end{equation}
This is like a $\mathcal{Q}$-map (\ref{Q}) but with general coefficients $(a,b,c)$. The proof is somewhat tedious and proceeds by considering every $\mathcal{L}_k$ separately.
\subsection{$\mathcal{P}^2$}
It is trivial to see that $\mathcal{P}^2(\Gamma) = \Gamma$ and that this is a generalized $\mathcal{Q}$ map with coefficients 
\begin{equation}
a = 1\qquad b = 0\qquad c = 0~.
\end{equation}
\subsection{$\mathcal{Q}^2$}
To reexpress $\mathcal{Q}^2$ we first calculate the various pieces,
\begin{eqnarray}
\bar{\mathcal{Q}}(\Gamma)_{\alpha\gamma} &=& \left[\frac{M-N-1}{N(N-1)}\right]\delta_{\alpha\gamma}\bar{\bar{\Gamma}} - \left[\frac{M-N-1}{N-1}\right]\bar{\Gamma}_{\alpha\gamma}~,\\
\bar{\bar{\mathcal{Q}}}(\Gamma) &=& \left[\frac{(M-N)(M-N-1)}{N(N-1)}\right]\bar{\bar{\Gamma}}~.
\end{eqnarray}
Substitute into Eq.~(\ref{Q}) leads once again to a generalized $\mathcal{Q}$ map with coefficients:
\begin{equation}
a = 1\qquad b = \frac{4N^2 + 2N - 4NM + M^2 -M}{N^2(N-1)^2}\qquad c = \frac{2N-M}{(N-1)^2}~.
\end{equation}
\subsection{$\mathcal{G}^\dagger\mathcal{G}$}
With the same strategy one finds on the basis of Eq.~(\ref{G_down}) and 
\begin{equation}
\bar{\mathcal{G}}(\Gamma)_{\alpha\gamma} = \frac{M-1}{N-1}\bar{\Gamma}_{\alpha\gamma}~,
\end{equation}
that substituting into (\ref{G_down}) leads to another generalized $\mathcal{Q}$ map with coefficients:
\begin{equation}
a = 4\qquad b = 0 \qquad c = \frac{2N - M - 2}{(N-1)^2}~.
\end{equation}
\subsection{$\mathcal{T}_1^\dagger \mathcal{T}_1$}
The needed terms are now:
\begin{eqnarray}
\bar{\mathcal{T}}_1\left(\Gamma\right)_{\alpha\beta;\gamma\delta} &=& (M-4)\Gamma_{\alpha\beta;\gamma\delta} + \left[\frac{M-N-2}{N(N-1)}\right]~(\delta_{\alpha\gamma}\delta_{\beta\delta} - \delta_{\alpha\delta}\delta_{\beta\gamma})\bar{\bar{\Gamma}}~,\\
&&- \left[\frac{M-N-2}{N-1}\right]\hat{A}\left[\delta_{\alpha\gamma}\bar{\Gamma}_{\beta\delta} - \delta_{\beta\gamma}\bar{\Gamma}_{\alpha\delta}-\delta_{\alpha\delta}\bar{\Gamma}_{\beta\gamma} + \delta_{\beta\delta}\bar{\Gamma}_{\alpha\gamma}\right]~,\\
\bar{\bar{\mathcal{T}}}_1\left(\Gamma\right)_{\alpha\gamma} &=& \left[\frac{(M-N-2)(M-N-1)}{N(N-1)}\right]\delta_{\alpha\gamma}\bar{\bar{\Gamma}} - \left[\frac{(M-3)(M-2N)}{N-1}\right]~\bar{\Gamma}_{\alpha\gamma}~,\\
\bar{\bar{\bar{\mathcal{T}}}}_1\left(\Gamma\right) &=& \left[\frac{(M-2)(M(M-1) - 3N(M-N))}{N(N-1)}\right]\bar{\bar{\Gamma}}~,
\end{eqnarray}
and substitution into Eq.~(\ref{T1_down}) leads to the coefficients:
\begin{eqnarray*}
a &=& M-4~,\\
b &=& \frac{M^3-6M^2N-3M^2+12MN^2+12MN+2M-18N^2-6N^3}{3N^2(N-1)^2}~,\\
c &=& -\frac{M^2 + 2N^2 - 4MN - M + 8N - 4}{2(N-1)^2}~.
\end{eqnarray*}
\subsection{$\mathcal{T}_2^\dagger \mathcal{T}_2$}
Finally, needed for the calculation of the last map are:
\begin{eqnarray}
\bar{\mathcal{T}}_2\left(\Gamma\right)_{\alpha\beta;\gamma\delta} &=& \frac{\bar{\bar{\Gamma}}}{N - 1}(\delta_{\alpha\gamma}\delta_{\beta\delta} - \delta_{\alpha\delta}\delta_{\beta\gamma}) + M~ \Gamma~,\\
&&- \left[\delta_{\alpha\gamma}\bar{\Gamma}_{\beta\delta} - \delta_{\beta\gamma}\bar{\Gamma}_{\alpha\delta} - \delta_{\alpha\delta}\bar{\Gamma}_{\beta\gamma} + \delta_{\beta\delta}\bar{\Gamma}_{\alpha\gamma}\right]~,\\
\tilde{\mathcal{T}}_2\left(\Gamma\right)_{\alpha\beta;\gamma\delta} &=& \frac{M - N}{N - 1}\bar{\Gamma}_{\beta\delta}\delta_{\alpha\gamma} + \delta_{\beta\delta}\bar{\Gamma}_{\alpha\gamma} - (M - 2)\Gamma_{\alpha\delta;\gamma\beta}~,\\
\tilde{\tilde{\mathcal{T}}}_2\left(\Gamma\right)_{\alpha\gamma}&=& \left[\frac{M(M - N) - (N - 1)(M - 2)}{N - 1}\right]\bar{\Gamma}_{\alpha\gamma} + \delta_{\alpha\gamma}\bar{\bar{\Gamma}}~,
\end{eqnarray}
which, when substituted into Eq.~(\ref{T2_down}) gives the following coefficients:
\begin{equation}
a = 5M - 8\qquad b = \frac{2}{N - 1}\qquad c = \frac{2N^2 + (M - 2)(4N-3) - M^2}{2(N - 1)^2}~.
\end{equation}
The overlap-matrix map is just the sum of the various terms obtained, and hence also a generalized $\mathcal{Q}$ map with rather complex coefficients.
\section{\label{inverse_overlapmatrix}Inverse of generalized $\mathcal{Q}$ map}
The inverse of a generalized $\mathcal{Q}$ map can be shown to be another generalized $\mathcal{Q}$ map. Consider for brevity the notation: 
\begin{equation}
\mathcal{Q}(a,b,c)(\Gamma) = Q~,
\end{equation}
then applying partial trace operations on Eq.~(\ref{Q_like}) leads to:
\begin{eqnarray}
\bar{\bar{\Gamma}} &=& \frac{\bar{\bar{Q}}}{a + M(M - 1)b - 2(M - 1)c}~,\\
\bar{\Gamma}_{\alpha\gamma} &=& \frac{1}{a - c(M - 2)}\left[\bar{Q}_{\alpha\gamma} - \frac{b(M - 1) - c}{a + M(M - 1)b - 2(M - 1)c}\delta_{\alpha\gamma}\bar{\bar{Q}}\right]~.
\end{eqnarray}
Upon substitution into Eq.~(\ref{Q_like}) and solving for $\Gamma$ one obtains,
\begin{equation}
\Gamma = \mathcal{Q}^{-1}(a,b,c)(Q) = \mathcal{Q}(a',b',c')(Q)~,
\end{equation}
where
\begin{eqnarray}
a' &=& \frac{1}{a}~,\\
b' &=& \frac{ba + bcM -2c^2}{a\left[c(M - 2) - a\right]\left[a + bM(M - 1) - 2c(M - 1)\right]}~,\\
c' &=& \frac{c}{a\left[c(M - 2) - a\right]}~.
\end{eqnarray}
These are important relations since they allow to evaluate the action of the inverse overlap matrix on a tp matrix as fast as a $\mathcal{Q}$ map. \emph{i.e.} at a computational cost which is negligible compared to the other matrix manipulations.
\bibliography{primal_dual.bib}
\end{document}